# Building and Evaluating a Realistic Virtual World for Large Scale Urban Exploration from 360° Videos

Mizuki Takenawa[1] · Naoki Sugimoto[2] · Leslie Wöhler[1] · Satoshi Ikehata[3] · Kiyoharu Aizawa[1]

**Abstract**
We propose to build realistic virtual worlds, called 360RVW, for large urban environments directly from 360° videos. We provide an interface for interactive exploration, where users can freely navigate via their own avatars. 360° videos record the entire environment of the shooting location simultaneously leading to highly realistic and immersive representations. Our system uses 360° videos recorded along streets and builds a 360RVW through four main operations: video segmentation by intersection detection, video completion to remove the videographer, semantic segmentation for virtual collision detection with the avatar, and projection onto a distorted sphere that moves along the camera trajectory following the avatar's movements. Our interface allows users to explore large urban environments by changing their walking direction at intersections or choosing a new location by clicking on a map. Even without a 3D model, the users can experience collision with buildings using metadata produced by semantic segmentation. Furthermore, we stream the 360° videos so users can directly access 360RVW via their web browser. We fully evaluate our system, including a perceptual experiment comparing our approach to previous exploratory interfaces. The results confirm the quality of our system, especially regarding the presence of users and the interactive exploration, making it most suitable for a virtual tour of urban environments.

**Keywords** Multimedia · Omnidirectional video · 360° video · Virtual reality

E-mail   Mizuki Takenawa
         takenawa@hal.t.u-tokyo.ac.jp
         Naoki Sugimoto
         naoki48916@gmail.com
         Leslie Wöhler
         woehler@hal.t.u-tokyo.ac.jp
         Satoshi Ikehata
         sikehata@nii.ac.jp
         Kiyoharu Aizawa
         aizawa@hal.t.u-tokyo.ac.jp

[1] Department of Information Communication and Engineering, The University of Tokyo, 7-3-1 Hongo, Bunkyo-ku, Tokyo 113-8656, Japan
[2] MMMakerSugi, Tokyo, Japan
[3] National Institute of Informatics, 2-1-2 Hitotsubashi, Chiyoda-ku, Tokyo, 101-0003, Japan

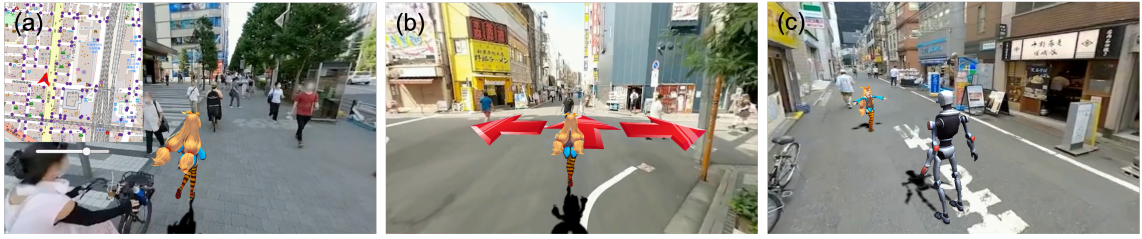

**Fig. 1** We present 360RVW: an approach for the creation of realistic and interactive virtual worlds directly from 360° videos. (a) Users can explore a photorealistic virtual world with moving people and vehicles via their avatar. (b) Users can select directions at each intersection to explore large areas. (c) Multiple users can explore the area simultaneously and interact with others via their avatar.

# 1 Introduction

Virtual worlds replicating real world locations are widely used across various applications ranging from entertainment to education. Typically, these virtual worlds reconstruct the real environment using 3D models, enabling users to navigate freely and interact through avatars. This allows users from all over the world to connect and participate in local events, such as celebrating New Year's Eve in the virtual Times Square [1], or Halloween in the Virtual Shibuya [2]. However, while the demand for creating virtual worlds that resemble the real world increases, creating these environments through manual modeling remains labor-intensive and time-consuming. In recent years, research has made progress in using deep learning techniques for automatic 3D model generation. However, these methods still heavily rely on the accuracy of training datasets and face significant challenges in terms of generalization and computational efficiency [3]. Practically, virtual exploration of large-scale areas is realized by directly utilizing images or videos such as Google Street View (GSV) [4] and MovieMap [5]. Unfortunately, this approach limits interactivity and makes it more difficult to create social experiences because users can only passively watch videos or images while moving along the trajectory of the camera.

In this paper, we create a photorealistic virtual world (360RVW) that combines the positive aspects of both 3D models and image-based approaches. Our virtual world allows users to freely explore a photorealistic virtual environment that replicates a wide real area and interact with other users via 3D avatars (Fig. 1 (a)). To achieve this, we first determine the real world coordinates for each video frame to estimate the camera trajectory and display 360° video frames corresponding to their respective positions. By rendering the surrounding scenery in our worlds solely through 360° videos, we avoid the manual creation of large and complex 3D models. To increase the interactivity of the system, we add a user's avatar to the scene which was shown as a promising direction for systems in small scale indoor environments [6]. In our system, the viewing camera follows behind the avatar on the estimated trajectory allowing users to freely explore large outdoor area and interact with others. As we do not use an explicit 3D model, we propose to use a semantic segmentation approach to implicitly model collisions and prevent the avatar to be rendered on the buildings. Based on these ideas, we combine previous multimedia technologies in novel and intuitive ways to create large-scale photorealistic environments with interactive controls from real world data.

Overall, creating our interactive virtual world requires only a set of 360° street view videos and is processed highly automatically, making it feasible even for novice users. The creation process involves the following steps shown in Fig. 2.

**(A) Video processing.** Following the methodology of MovieMap [5], we first detect intersections in the videos and match the frames to real world coordinates. This way, we can easily choose the correct frame for rendering even when the avatar is moved between areas that originally did not belong to the same video.

**(B) Video completion.** As we use real world 360° videos, there is always a videographer in them. Therefore, we removethe videographer from the videos by video completion to prevent it from being visible in the virtual environment. For highly accurate completion, we apply rotation to target videos. Rotation is a manipulation unique to 360° images that can reduce distortions present at the poles of the equirectangular projection (ERP) images by rotating the image content to the center before further processing.

**(C) Virtual collision detection.** To enhance the realism of the virtual environment, we implement collision detection between the avatar and buildings by using semantic segmentation to identify walkable roads.

**(D) Dynamic projection.** We project the 360° frames onto a sphere that moves along the recorded camera trajectory following the avatar's movements. When walking in a city street in our virtual world, the avatar's movement and video playback are synchronized to give the user the sensation of walking through the city.

**(E) Interface.** We implement an interface to allow users to explore large urban areas by changing the avatar's walking direction at intersections (Fig. 1 (b)). Users can also quickly teleport to a new location by choosing a location on the map. Furthermore, we use video streaming that reduces data loading to a minimum for quick rendering and makes it easy to share the virtual worlds. The overview of these proposed method is shown in the supplementary video, Online Resource 1.

To evaluate our virtual worlds and interface, we conducted user experiments comparing 360RVW to GSV [4] and MovieMap [5] focusing on the perceived exploration experience and usability of the interface. We also evaluated virtual collision detection, and video completion techniques comparing with and without rotation.

In summary, our contributions are as follows.
- We propose a simple yet effective approach to create large-scale interactive virtual worlds directly from 360° videos, that contributes to applications of 360° videos.
- We propose rotation to improve the completion of 360° videos and quantitatively evaluate our proposed techniques to validate their quality.
- We realize virtual collision detection between avatars and buildings via semantic segmentation. The results of the user study suggest that the collision detection is sufficiently accurate for users to perceive a three-dimensional effect without discomfort.
- We evaluated our interface against GSV and MovieMap in the task of exploring along specific routes. Our interface received better ratings compared to MovieMap and GSV.

In our previous work, we demonstrated 360RVW [7], but the exploration was limited to a small area and the system lacks functions needed for large urban areas. In this paper, we show a comprehensive 360RVW for large scale urban exploration, and quantitively and qualitatively evaluate the exploration by avatar via a user study.

# 2 Related works

Urban visualization has been approached in various ways, including the use of images, videos or novel view synthesis such as NeRF [8] and Gaussian Splatting [9].

## 2.1 Virtual exploration via 360° images

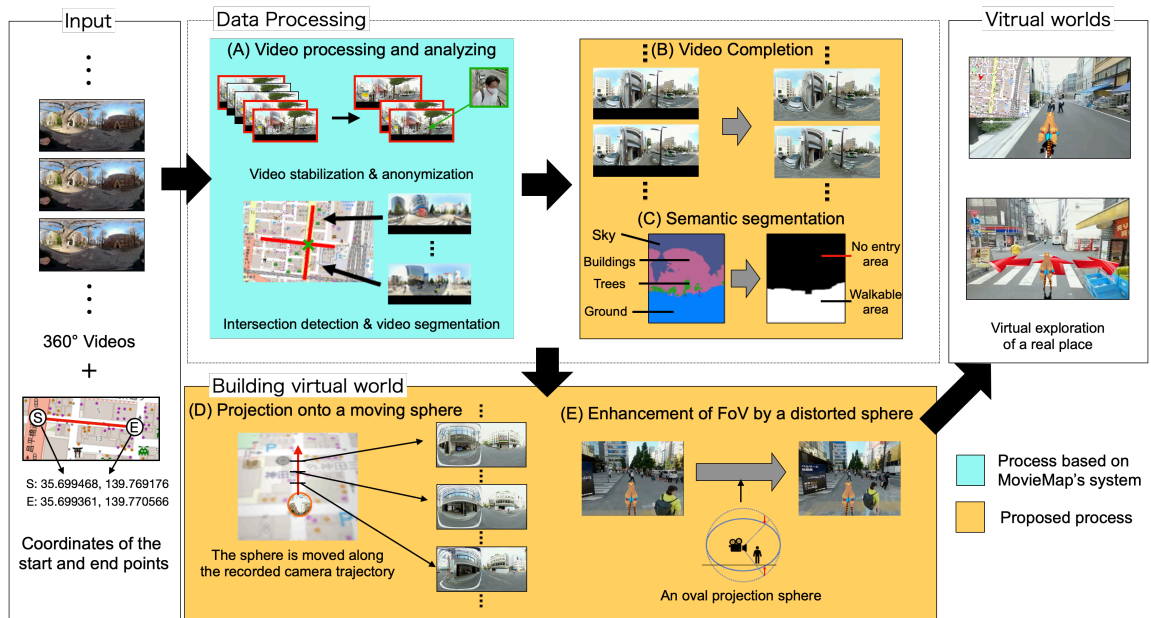

**Fig. 2** Outline of our system flow. The process colored by blue is based on MovieMap's system [5] and that colored by orange is our proposal.

GSV and Mapillary [10] are applications that allow users to view information about a location while moving through the city, by associating 360° images captured on-site with an abstracted map interface. These approaches were adapted by various methods which propose to integrate social media features. Social Street View (SSV) [11] proposes an algorithm to display geo-tagged social media on 360° images of the corresponding location. Similar to GSV, SSV uses sparsely positioned 360° images, hence users cannot walk seamlessly through the streets, and interactions with other users are not possible. Geollery [12] aimed to solve this problem and proposed a method to create a 3D map of an area selected by the user based on GSV images and adds social media in the 3D space. In the world made by Geollery, users move inside a 3D city where GSV images are used as textures, while also engaging in real-time interactions with other users via their avatars. Because Geollery switches the 360° image used as the texture depending on the user's position to accommodate changes in viewpoint, the texture of the cityscape changes significantly and this leads to the reduction of realism.

Avatar360 [6] proposed a system that allows users to explore shooting locations via their avatar while switching between multiple 360° images. The result of this research implies that users prefer using avatars than click and drag to explore, and that the camera automatically follows the avatar is the best viewpoint control in 360° images. Our work improves over Avatar360 by enabling smooth movements through videos instead of movements between images. Furthermore, our approach is not limited to indoor scenes and can realize large virtual worlds.

## 2.2 Virtual exploration via 360° videos

Aspen Movie-Map [13] is the seminal work of interactive videos that represents a virtual world where users can explore movies taken along the streets in the city. The work is very early and depends on analog technology, such as film cameras and analog optical disks, that must have required huge manual operations. No other Movie-Map has been created before Sugimoto et al. revisited MovieMap by today's digital image processing techniques [5]. With the availability of inexpensive and small 360° cameras, MovieMap becomes practical. In MovieMap, the detection of intersections and the segmentation of videos are automated. Unlike GSV, users can feel the atmosphere of the recorded area through moving pedestrians and vehicles. However, MovieMap only provides an interface in which individual users watch the videos and do not provide interaction with other users. In contrast, Tourgether360 [14] created a simplified virtual world from a single 360° video, enabling interaction with other users. In Tourgether360, the camera trajectory is visualized, and the current viewing position of others is displayed by spherical avatars in the shape of eyes, so that users can see which parts of the video others are being watched. Users can move only on the camera trajectory and cannot freely explore the environment.

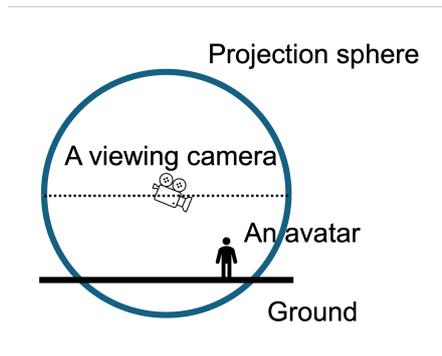

**Fig. 3** A component of 360RVW. 360RVW has no city 3D model and is mainly composed of projection spheres, the ground plane and an avatar.

## 2.3 Recent 3D techniques

Techniques such as NeRF [8] and Gaussian Splatting [9] enable novel view synthesis from a sufficient number of input images of static scenes, and are therefore also utilized for 3D urban visualization [15,16]. Although these two approaches differ in the output produced by their respective CNNs, their advantages and disadvantages are largely similar. Both methods can generate highly realistic reproductions of the input scenes and allow for free manipulation of the camera viewpoint. However, they intrinsically deal with static objects and cannot reproduce a scene with dynamic objects such as moving cars and walking people, as is the case outdoors in the real world.

The 360RVW approach constructs virtual environments that replicate the real world by directly utilizing video footage. Although this method fixes the camera viewpoint in the virtual space to the position of the camera during filming, it can reproduce lively moving scenes of the streets. In addition, it only requires capturing videos of the target area. Expanding the coverage of the target region can be easily achieved by simply acquiring additional videos of the new areas to be included.

## 2.4 Video completion

Video completion is the task of removing unintended objects or patterns in the video from all frames. In this task, mask sequences are provided that indicate areas for completion. Therefore, temporal consistency with adjacent frames must be considered in addition to spatial consistency within a single frame. In deep video completion models, there are primarily two approaches, flow-guided-based methods [17-19] and transformer-based methods [20-22]. There is also an approach by combination of the two [23, 24].

While some previous work exists for 360° images, to the best of our knowledge, there are no completion methods for 360° videos. Due to the lack of datasets, all existing deep video completion models have focused on perspective projection videos. In this study, we fully utilize the conventional completion techniques optimized for the perspective projection and apply them to ERP format frames of 360° videos by performing rotation to the frames.

### 2.5 Semantic segmentation

Semantic segmentation is one of the major computer vision tasks that predicts the classes of pixels among a set of predefined classes [25]. There are semantic segmentation models [26], proposed for 360° images captured on outdoor roads. In our study, we apply semantic segmentation to each frame of 360° videos to acquire its road region, and we use the region as its meta-data of the frame. We use this meta-data to determine where the avatars can move in a virtual world. Restricting the avatar's movement within the road regions eventually leads to virtual collision detection when the avatar moves beyond the border between the road and the building. We utilize Orhan's semantic segmentation model [26] in our system. Based on its output, we define only *Ground*, *Road*, *Sidewalk*, and *Parking* classes under the *Flat* category, as well as the *Person* class under the *Person* category in Orhan's model, as navigable for avatars. All other classes are treated as non-navigable.

Ordinally, model-based virtual worlds[11, 12] enable rich interactions between avatars and the space, however, visualization is not photorealistic. In contrast, image- or video-based virtual environments can provide photorealistic visuals but offer limited interaction. Our approach employs videos, and their data created from video processing for rendering the environment while integrating avatars within the space, thereby enabling user interaction in a hybrid manner.

## 3 Building 360RVW

We aim to create interactive virtual worlds using 360° videos captured by users, developing a system that does not require specialized knowledge of 3D modeling. Therefore, as input, our system only requires 360° videos captured along city streets and the start and end coordinate of each video. Instead of using 3D textured models of cityscapes, we simply project the videos onto a spherical surface to create the illusion of a city's presence as shown in Fig. 3. We only use minimal 3D models, such as projection spheres, a ground plane, and an avatar in the virtual world. The avatar can move freely and the camera exists in the viewing center of the sphere and is always facing towards the avatar. We utilize these 3D models for visualization of other users and interaction with not only users but also environments. To realize our system, we follow the video processing steps described by MovieMap [5] and implement the virtual world with Unity [27]. The pipeline our system is visualized in Fig. 2 Outline of our system flow. The process colored by blue is based on MovieMap's system [5] and that colored by orange is our proposal.. We simplify the explanation of 360° video processing and analyzing that are common with MovieMap.

### 3.1 360° Video Processing and analyzing

Following the video processing of MovieMap [5], we first perform video stabilization [28] to reduce camera shaking due to walking while recording. Then, we anonymize the faces by blurring to ensure the privacy of pedestrians in the videos. Finally, we detect intersections and segment the videos at each intersection. For the intersection detection, we apply OpenVSLAM [29] to the videos and provide start and end coordinates to estimate the camera position for each frame in every video. This way, we assign the coordinates to all frames and map them to a shared space. From this spatial information, we find pairs of frames that are intersections from different videos and segment the videos at intersections.

Thus, we obtain stabilized, anonymized video segments with aligned spatial position information in each frame.

## 3.2 Video Completion

In this study, we use 360° videos captured in the real world, therefore it's inevitable that the videographer is visible. When exploring the 360° videos using a camera that follows the avatar with its line of sight as shown in Fig. 3, there may be a chance for the camera view to face the ground, causing the videographer to overlap with the avatar. Therefore, we perform video completion to remove the videographer from the 360° videos applying the existing models to 360° images in the format of ERP image. In our system, we use videos captured with a 360° camera placed above the videographers' heads. Therefore, in an ERP format of 360° videos, the videographer largely appears at the bottom of the image, which often leads to failures when applying previous deep video completion models because they are trained on perspective projection video datasets [31]. However, ERP format of 360° images and videos has distortion, especially large in the bottom and top area. Due to this distortion, previous deep video completion models tend to fail to provide a plausible completion of 360° videos.

Therefore, as shown in Fig. 4, we propose to utilize the spherical properties of 360° videos by rotating each frame to move the videographer to the center, and after removal and completion of the yellow region, rotate them back to the original projection. Centering the completion region has two advantages. One is to reduce the distortion which is strongest at the top and bottom of 360° images, and the other is to reduce the area of the completion region. Removing the distortion is especially useful, as it preserves straight-line patterns like road markings and building corners. After rotating, the video completion can be handled by conventional models trained on perspective projection videos. We conducted the quantitative and qualitative experiment for existing video completion models to evaluate effect of the rotation. The results of the experiment are given in Online Resource 2.

## 3.3 Semantic segmentation for collision detection

As we only use 360° videos without 3D building models for rendering a cityscape, including collision detection is not straightforward. Without collision detection, users can move their avatar freely over buildings, resulting in appearances that are not possible in reality. To avoid damaging the realism of our virtual world without introducing 3D models, we set up collision detection based on image content.

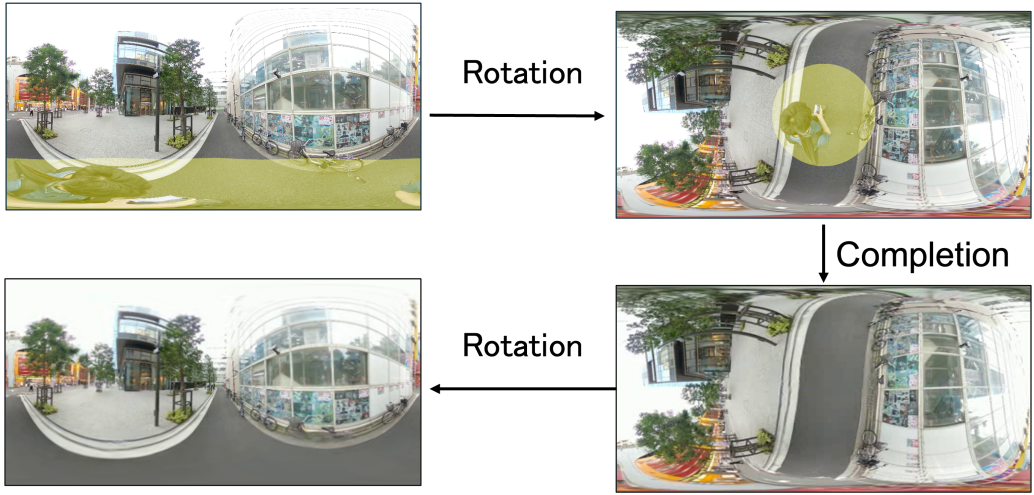

**Fig. 4**  A method for high quality completion of 360° videos using rotation.

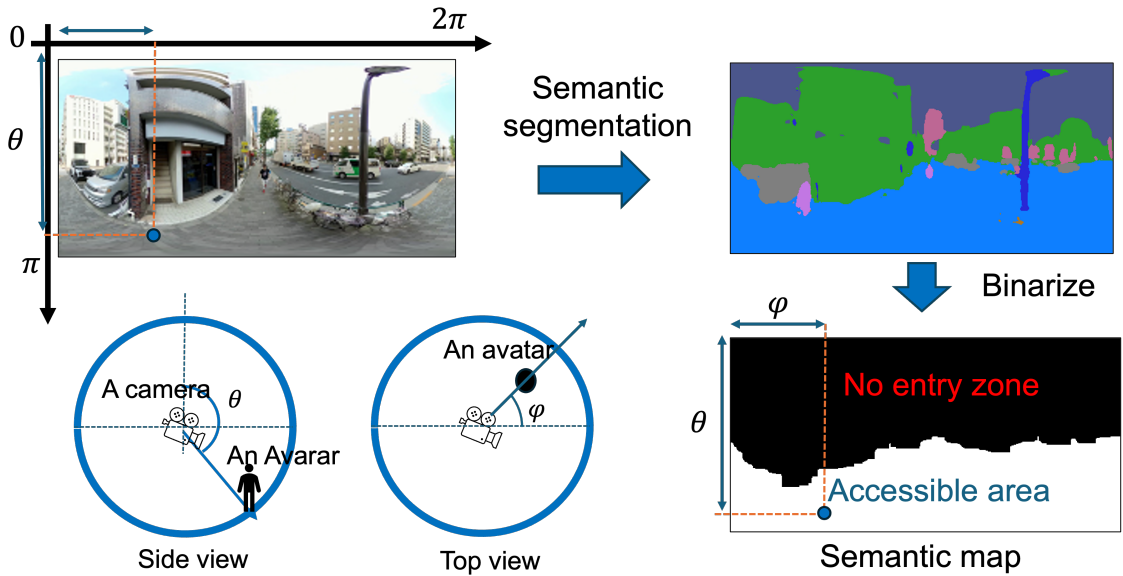

**Fig. 5**  Making segmentation maps for collision detection and how collision detection is performed.

Towards this goal, we propose to use semantic segmentation. As semantic segmentation is used to predict classes for each image pixel, we can utilize this information to estimate the image positions of buildings and streets. This allows us to restrict the avatar's movement to the road regions and eventually leads to virtual collision detection when the avatar tries to move beyond the border between the road and the building. The creation process is visualized in Fig. 5. First, we apply semantic segmentation [26] to each frame of the videos to estimate the road region using a previous approach designed for 360° videos of cities. Afterwards, we binarize the results to make semantic maps, only differentiating between walkable (streets) and non-walkable areas (buildings). To realize collision detection, we calculate the pixels on which the avatar is standing. We can get the position of these pixels by simply using latitude and longitude of the avatar's foot because the vertical and horizontal sides of ERP images respectively correspond to latitude and longitude (bottom left of Fig. 5). Thereby, we can easily realize collision detection by preventing avatar's position from invading non-walkable areas (bottom right of Fig. 5). We also allow avatars to overlap other regions, such as pedestrians, to facilitate a more comfortable exploration experience. Semantic segmentation assigns only one class to one pixel, and we carefully choose classes, that are road and pedestrians to be included in the movable regions of avatars. This method determines the avatar's position on the ERP based on latitude and longitude, and thus it is not significantly affected by the binary map's resolution, the resolution of the video projected onto the sphere, or the size of the projection sphere. Furthermore, since one binary map corresponds to its corresponding frame, it is updated to the locations where collision detection is applied.

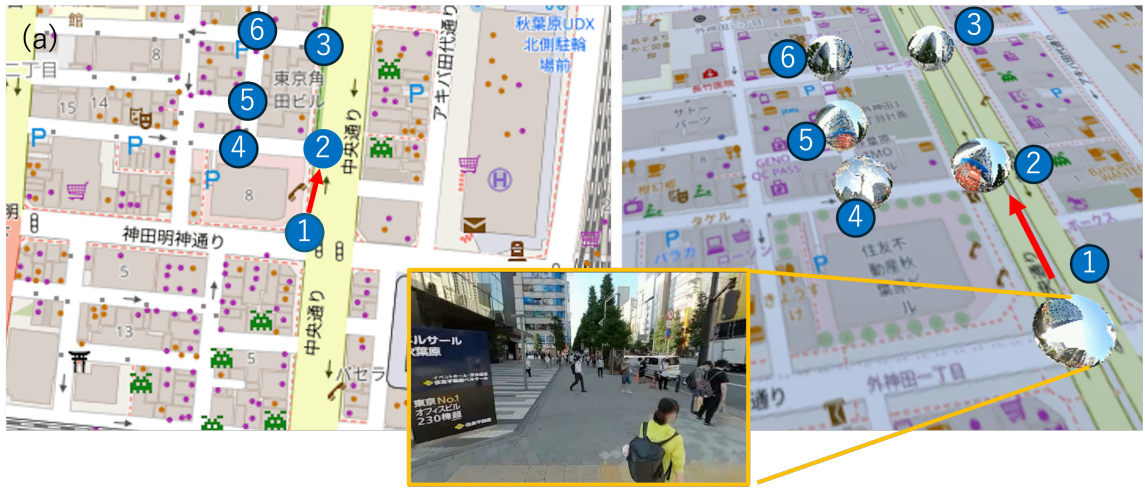

**Fig. 6** The positions of the projection spheres in 360RVW. The projection spheres in virtual worlds are placed at intersections that maintain the same spatial relationship as in real space.

To realize large scale areas and fast streaming for 360RVW, we represent the semantic maps in JSON format to easily access the pixels in each frame. We apply compression to the semantic map. Since semantic maps have only black and white pixels, we apply a general run length encoding for compression. Thanks to this, we can drastically reduce the file size of the semantic map, for example, semantic maps in the entire Akihabara area is reduced to 1 percent of the initial size.

### 3.4 Projection onto a moving sphere

In 360RVW, as the user moves the avatar forward, the video segment projected onto the sphere shows the changing scenery, and the sphere itself moves forward in the same 3D space as the avatar. They can explore a large area as our system changes from one segmented video to another at each intersection.

Fig. 6 shows an example of the projection spheres at their initial locations. The projection sphere moves along a street from one intersection to another in the virtual space. When the avatar moves to a particular direction, the corresponding video segment is played and projected onto the sphere that moves on the camera trajectory estimated by vSLAM. When the user walks on the road, the projected scenery has no artifacts as the camera is fixed to the actual captured position and shows the original video frame. Just before approaching an intersection, the next possible video segments are pre-loaded in advance and projected onto the projection spheres, and the user selects one of them.

### 3.5 Enhance FoV by a distorted sphere for the projection

The field of View (FoV) of 360° images is one of main parameters on user's experience, and it was concluded that a FoV close to 110° is the best for users when viewing 360° videos on 2D displays [28].

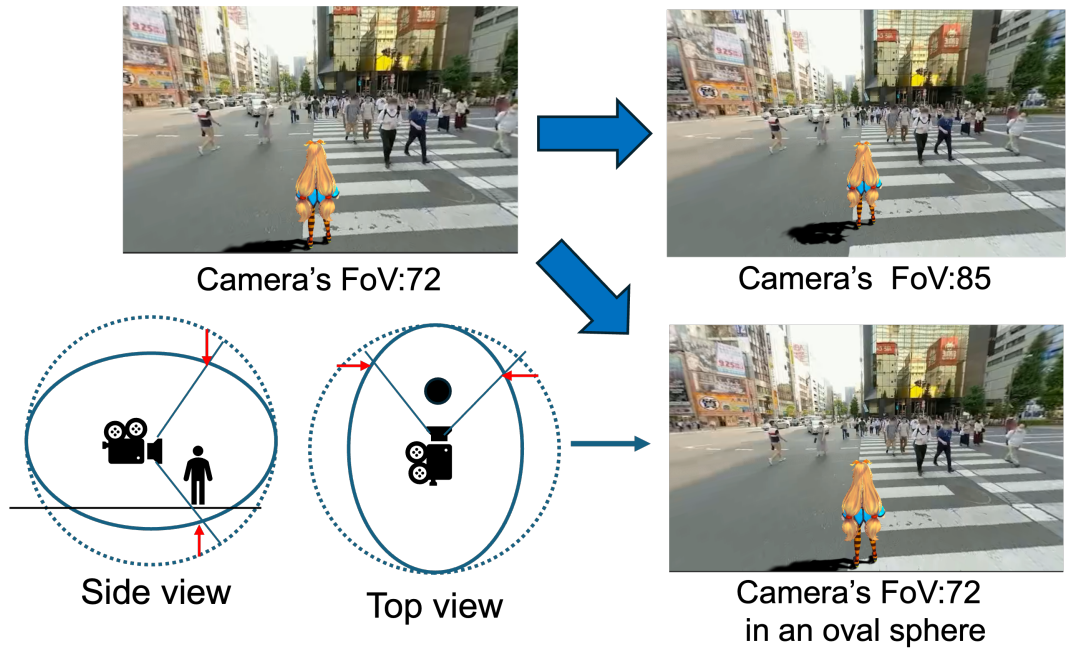

**Fig. 7** The effect of projecting 360° videos into an oval sphere surface.

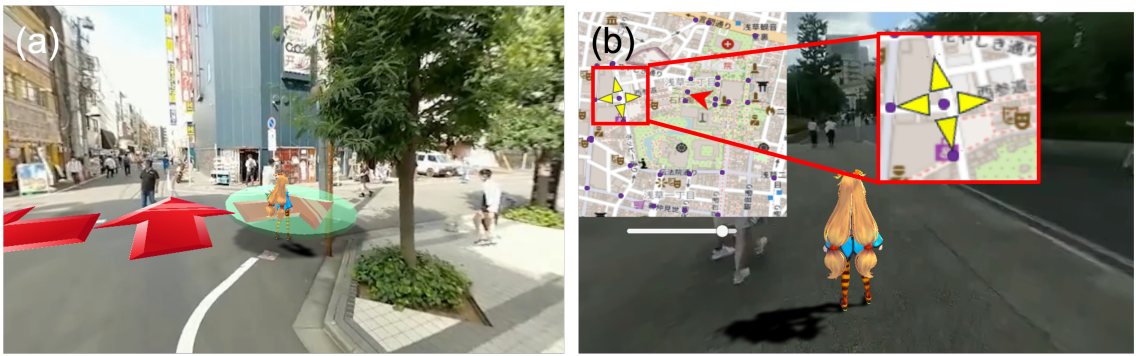

**Fig. 8** Choosing the direction. (a) Users can select the direction by overlapping an avatar with a red arrow at intersections. (b) Users can warp to a remote location by clicking on the yellow arrow on the map.

Since the default value of the FoV in Unity is 60°, we enhance the FoV by altering the projection surface. Specifically, we propose replacing the conventional spherical projection used for 360° videos with an oval projection that resembles the shape of a rugby ball. This approach for enlarging the FoV is distinct from merely increasing the FoV settings of the camera in virtual worlds for the user's view. Fig. 7 illustrates the differences between views in our method and the one obtained by simply enlarging the camera's FoV. When the camera's FoV is increased, avatars and backgrounds appear more distant. On the other hand, projection onto an oval sphere can extend FoV without changing the avatar's position in the screen. This approach inherently results in significant distortion near the poles due to the oval shape. Since the vertical field of view is typically limited and the poles are outside the viewer's focus, the distortion is not very visible in the user experiences.

# 4 An interface for exploring 360RVW

We propose an interface for exploration of a virtual world created by 360° videos. Users can operate their own avatar with a PC keyboard or a gamepad.

Fig. 1 provides an overview of the interface, where users can view their current location on the map and navigate through a realistic virtual world. At intersections, users have the option to select their preferred direction. Additionally, the map displayed on the screen can be toggled between hidden and visible modes, enabling users to check their location at any time they want while exploring on the screen, which usually has a wide view.

At intersections, users can choose their direction by either positioning their avatar in line with a red arrow or using keyboard direction keys (see Fig. 8 (a)). Additionally, our virtual world is constructed from 360° videos recorded by walking the roads in both directions. This setup allows users to switch to videos showing the opposite direction by selecting the arrow that points backwards. Using bidirectional videos on one path can avoid playing the video backwards when users go in the opposite direction. When switching to the opposite direction, the system locates the frame in reverse direction footage with the closest coordinate to the current position and playback from that frame, thereby minimizing discontinuities in position. Users can also navigate to different exploration areas by choosing intersections and yellow arrows displayed on the map, giving them the freedom to explore any area they wish (Fig. 8 (b)). Transition of videos shows the discontinuity of the backgrounds in the different videos and results in disrupting user's sense of immersion. To maintain user's immersive experience, in this system, when changing videos by selecting the arrows, the display fades to black during the switch and after completing transition, a new video is displayed by a fade-in effect.

360RVW is a virtual world characterized by its three-dimensional scope. In this environment, users can interact with each other (Fig. 1 (c)), and also actively engage with objects in their surroundings. For instance, they could leave personal records within the world. Examples of such interaction include writing reviews about shops visible in 360° views or sharing messages of impressions at tourist spots they have visited.

The results in this paper were created from 360° videos recorded by Insta360 or GoPro Max cameras. The resolution of the video projected onto the sphere is 2K (1920×960), and the system operates smoothly with the video encoded in mp4 at a bitrate of 13Mbps on average. The system also supports different cameras and resolutions.

# 5 Evaluation of 360RVW: User study

## 5.1 Experimental setting

To assess the usability of our 360RVW interface, the sense of spatial presence, and advantage of navigating with an avatar, we designed an experiment simulating virtual tourism. In this experiment, we compare 360RVW to MovieMap and GSV by assessing the sense of presence, spatial cognition, and usability. The overview of differences among GSV, MovieMap, and 360RVW is shown in Table. 1.

**Table. 1** Comparison among three interfaces used in a user study.

|  | GSV [4] | MovieMap [5] | 360RVW |
|---|---|---|---|
| Media | Images | Videos | Videos and avatar |
| Operation | Mouse | Mouse | Keyboard or gamepad |
| Exploration | Select next locations | Select directions at intersections | Move an avatar in the world |

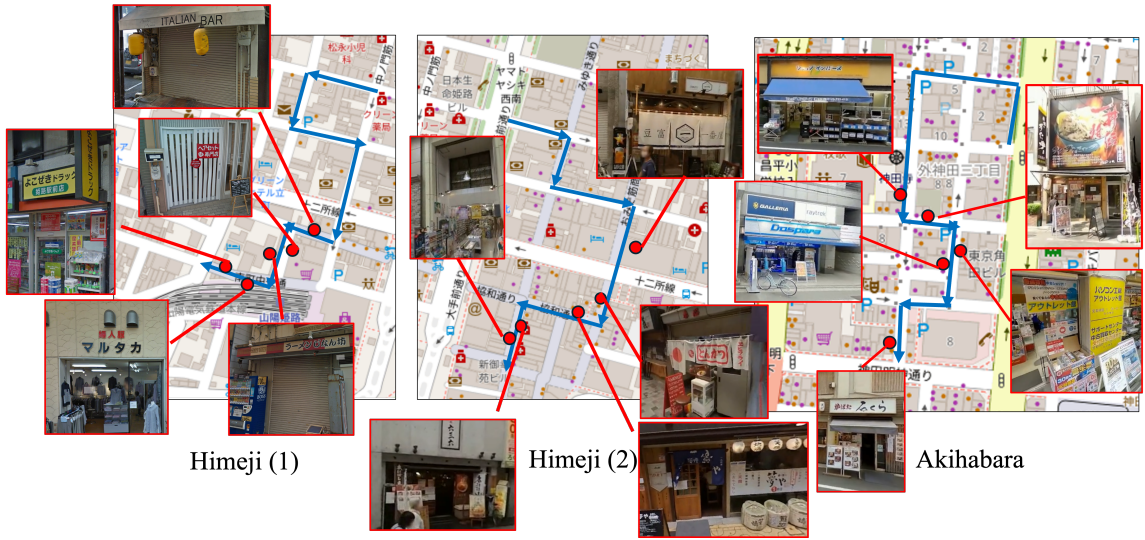

**Fig. 9**  An overview of the trial routes: There are two routes in Himeji and one in Akihabara. Blue arrows indicate the designated paths that participants will explore, while red dots indicate the landmarks that they will be asked to locate.

### 5.1.1 Experimental procedure

In this experiment, participants engaged in a virtual tour by exploring predefined routes and observing landmarks along the way. We used three different routes located in Japanese cities, Akihabara and Himeji, which included crowded shopping streets and roads. For each route, five distinctive landmarks, such as restaurants and shops with unique exteriors, were selected as stimuli (Fig. 9). Each participant experienced all three routes and all three exploration systems on a PC: GSV, MovieMap, and 360RVW. The order of routes and systems was randomized for each participant. Participants were informed in advance about the locations and appearances of the landmarks and were instructed to identify them while navigating the route from start to finish without a time limit. All three interfaces enable users to go back or go to specified location on the map, but participants were told not to use these functions in this experiment to prevent participants from deviating from designated route and getting lost.

Before starting the trials, we explained the task, obtained informed consent, and collected participant information. We also provided instructions on how to operate each exploration system: a gamepad was used for 360RVW, while a mouse was used for GSV and MovieMap. During the experiment, participants explored one route at a time, using one of the systems. After each exploration, they completed a questionnaire assessing their memory of the landmarks, their spatial presence, and their overall experience. Each system was used exactly once per participant. After completing all trials, we conducted interviews to gather additional feedback and then thanked participants and provided their reward.

**Table. 2** The results of preliminary questionnaires for participants. They were asked about their experience with virtual tours and games on the scale of 1(low experience) to 7 (high experience).

| Category | Summary |
| --- | --- |
| Age | 23.8 $\pm$ 3.06 |
| Gender | 19 (Male: 9, Female: 10) |
| Experience with first-person games (/7) | 3.74 $\pm$ 1.91 |
| Experience with third-person games (/7) | 4.16 $\pm$ 2.22 |
| Experience with virtual tours (/7) | 3.53 $\pm$ 1.93 |

### 5.1.2 Participants

The participants' backgrounds in this experiment were summarized in Table. 2. We recruited 19 participants on the web, consisting of 9 males and 10 females, all aged between 20 and 30 (M=23.8, SD=3.06). All participants regularly used a PC and were familiar with their operation. Participants reported varying levels of gaming and virtual tour experience on a 7-point scale, with the mean experience for first-person games being 3.74 (SD=1.91), for third-person games being 4.16 (SD=2.22), and for virtual tours being 3.53 (SD=1.93). Notably, 4 participants frequently played first-person view games (scores of 6 or 7), and 8 frequently played third person view games. Although 9 participants stated they have sufficient experience of using virtual tours including GSV (scores over 4), only 2 of them were heavy users of such services (scores of 6 or 7). Each participant spent 1.5 hours performing the entire series of tasks and received an Amazon gift card of 2000 yen for the participation. This experiment was approved by our university's ethics committee.

### 5.1.3 Measures

We aimed to assess the sense of presence and spatial cognition of participants as well as the usability of each system. Towards this end, we employ different questionnaires and tasks.

The presence of participants was measured using the Igroup Presence Questionnaire (IPQ) [32], which consists of four factors: overall sense of presence (PRES), sense of presence in a virtual world (SP), awareness of the outside world (INV) and sense of reality (REAL), each of which is scored on a Likert-scale from 1 to 7. The feeling of exploration and usability of the systems were evaluated by our original questions shown in Table. 3.

We showed the participants images of various landmarks, asking them to confirm whether these appeared along the route. The possible answers were "existed" or "didn't exist". We also showed intersections images, asking them to indicate the directions they took following the protocol of a previous study on spatial cognition in virtual worlds [33]. The possible responses were "forward", "left", "right", or "the intersection didn't exist".

**Table. 3** The items of questionnaire for the feeling of exploration. We use original questions using a 7-point Likert scale (7- fully agree, 1- fully disagree). Q2 means that the score is higher, participants feel they overlook landmarks more easily.

| ID | Question |
| --- | --- |
| Q1 | It was easy to find a landmark using the interface. |
| Q2 | You thought you were going to miss a landmark. |
| Q3 | You felt like you were actually walking around the city. |
| Q4 | You remembered the local scene and atmosphere. |
| Q5 | It was easy to walk along the specified route. |
| Q6 | You could explore in your own pace. |
| Q7 | You could freely expore the route |

Table. 4  The scores of the spatial cognition questionnaire. The number of participants who are highly experienced in first-person games is 4 and that of highly experienced in third-person games is 9. The scores of highly experienced in virtual tours is not displayed because there were only two.

|  | GSV | MovieMap | 360RVW |
|---|---|---|---|
| All | 11.37 ± 2.41 | 11.74 ± 2.16 | 11.47 ± 2.12 |
| Highly experienced in first-person games | 8.75 ± 3.95 | 10.75 ± 2.99 | 11.25 ± 3.5 |
| Highly experienced in third-person games | 11.63 ± 2.13 | 11.75 ± 1.75 | 11.75 ± 2.38 |

Table. 5  Distribution of responses to questions about the overall interface. The value represents the number of times selected.

| Questions | GSV | MovieMap | 360RVW |
|---|---|---|---|
| The interface easist to use | 6 | 9 | 4 |
| The interface most able to remember the scene | 9 | 5 | 5 |
| The interface most suitable for virtual tours | 3 | 5 | 11 |

After all trials, we conducted an additional questionnaire and interview to collect feedback on the overall experiment. We asked participants which interface is the easiest to use, which interface helped them remember the scenery the best, and which interface is most suitable for virtual tours. During interviews, we asked participants to explain their scores and discuss how the systems used in this experiment differed from each other.

## 5.2 Results and discussion

The results are shown in Table. 4, Table. 5 and Fig. 10, Fig. 11. Based on the results and the content of the interviews, we discuss the experimental outcomes. In this section, we refer to the 19 participants as P1, P2, and so forth.

### 5.2.1 Evaluation of the spatial cognition

To analyze the spatial cognition, we first calculate the average of correct answers shown in Table. 4. We calculated participants' scores by totaling correct responses with the maximum score being 18. As shown in Table. 4, The score that participants gave for 360RVW (M=11.47, SD=2.41) is similar to that for GSV (M=11.37, SD=2.16) and MovieMap (M=11.74, SD=2.12). In each interface, participants were on average able to correctly answer 11 of the 18 questions for all systems. From the point of view of spatial cognition, the three systems can be evaluated as statistically not different, despite their differences in viewing experience. Even participants with extensive experience in first-person games, third-person games did not necessarily achieve higher scores. This suggests that familiarity with similar interfaces or prior virtual tour experiences may not have a significant impact on memory of the exploration locations.

### 5.2.2 Evaluation of IPQ questionnaires

In contrast to the result of the spatial cognition, the three interfaces tend to differ in IPQ questionnaires, see Fig. 10. Among them, we detect significant differences in PRES and SP. In PRES, we detect a large difference between MovieMap (M=3.47, SD=1.58) and GSV (M=2.26, SD=1.32). 360RVW (M=3.21, SD=1.03) has a higher score than GSV, but there is no significant difference. In SP, significant differences exist between MovieMap (M=3.01, SD=1.19) and GSV (M=2.25, SD=0.74), and between 360RVW (M=3.18, SD=0.82) and GSV. The results in INV and REAL don't show the significant differences, but they suggest that 360RVW tends to be evaluated higher than GSV and comparable to MovieMap. In all cases, our findings show that 360RVW and MovieMap significantly outperform GSV. While 360RVW offers a third-person perspective and MovieMap provides a first-person view, their differing viewing experiences are compatible and both technologies consistently exceed the performance of GSV across all evaluated criteria. While GSV switches between still images to simulate movement, both MovieMap and 360RVW utilize video sequences, resulting in continuous scene transitions. This continuity is likely to contribute to a stronger sense of being present and moving within the environment.

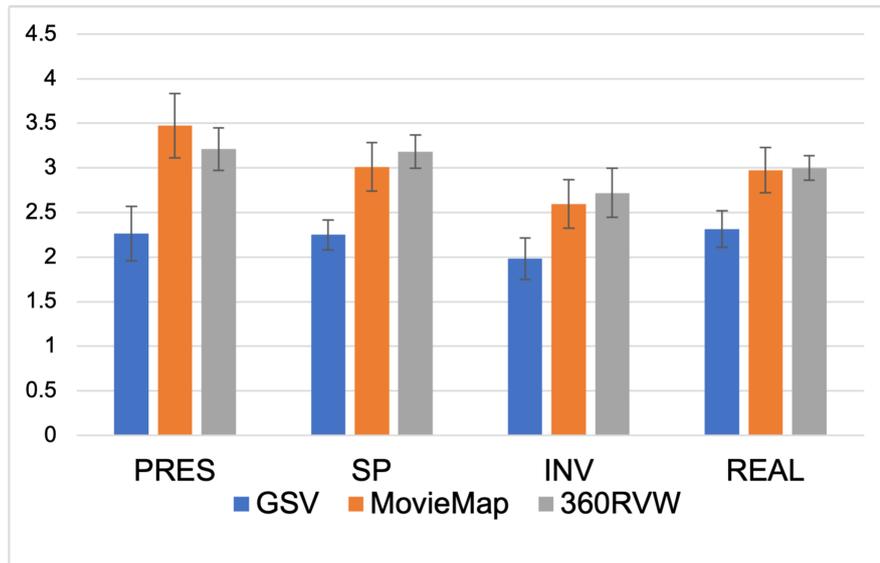

**Fig. 10** The score of Igroup Presence Questionnaire (IPQ) questionnaires.

### 5.2.3 Evaluation of overall experiences

Fig. 11 shows the results for the sense of exploration, based on our questionnaire (see Table. 3). The chart reveals that 360RVW consistently receives higher ratings than both MovieMap and GSV on nearly all items and thus secures a better overall score than MovieMap. 360RVW outperforms GSV and MovieMap especially in Q5, Q6 and Q7, that relate to exploring the specified route. It is suggested that 360RVW has a particular strength in allowing users to freely explore routes at their own desired pace.

In fact, MovieMap was criticized by the participants for its cumbersome adjustment of the video playback speed, which requires the user to stop moving in order to look around ([P3, P8, P13, P17]). Such inconvenience of adjustment may also be related to the ease of missing landmarks. For GSV, it was pointed out that users had difficulty navigating because they had to search for the locations of the placed 360° images ([P1, P11, P17, P19]). Some participants ([P7, P8, P12, P18, P19]) highly rated MovieMap for its strong sense of reality in exploration, attributing this to the first-person viewpoint and minimal controls which allowed them to focus more on the environment. Conversely, they rate 360RVW lower because operating an avatar with a gamepad was seen as laborious and detracting from the experience. However, this aspect of operation was valued by participants ([P3, P4, P5, P6, P14, P16]), who appreciated the proactive nature of virtual exploration, noting that the gamepad controls enhanced their sense of freedom to explore.

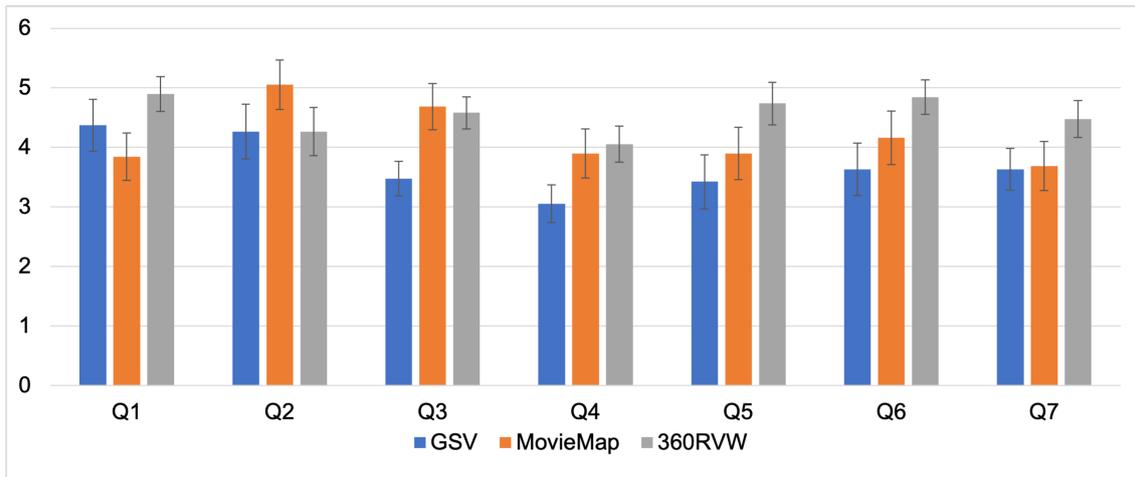

**Fig. 11** The score of each question for exploration.

Table. 5 presents the results concerning the overall questions about the systems. Here, MovieMap was considered the easiest for use, while 360RVW was favored for its suitability in virtual exploration. GSV was memorable for location exploration; however, considering Fig. 11 suggests, GSV is not the best choice for virtual exploration.

Participants reported that using 360RVW made them feel as if they were walking through the actual location via their avatars ([P2, P7, P9]). In particular, P16 mentioned that they felt the need to avoid colliding with people in the video. This sense of awareness comes from experiencing the environment through the avatar, making it a unique experience characteristic of 360RVW.

During the interviews conducted in this experiment, participants were asked about their awareness of collisions between the avatar and buildings in 360RVW. Six participants ([P4, P5, P8, P11, P13, P14]) reported noticing collision detection. Among them, five stated that the collision detection accurately matched the buildings' positions in the video and felt natural. One participant noted that the avatar's collision with the building as slightly off the wall, but this did not bother him because it is a common occurrence in video game experiences. Totally, the findings suggest that employing a semantic map for building collision detection could be sufficiently precise for virtual exploration.

## 6 Limitation

One limitation of this study is that the user evaluation was conducted with individual participants, and did not assess the system's performance or user experience in a multi-user virtual tour setting. Although the system is designed to support multiple users, as shown in Figure 1 (c), the effectiveness and usability of this functionality remain unverified. Future research should include evaluations with multiple users to explore the potential benefits and challenges of collaborative virtual tours. Another limitation is that the user study primarily relied on the small number of qualitative feedback. Future work should conduct the same experiment to get larger data about the differences of the interface and design new experiment featuring a specific function and incorporate quantitative measures, such as the number of collisions with segment maps and arrow presses to change videos to complement the findings.

The system is scalable and applicable to any area as long as videos are captured. The limitation of this system is to largely depend on the videos recorded manually. Urban scenes change from time to time, and updating the system needs extra efforts in video capture. Variations in shooting methods and walking speeds by different individuals directly affect the user experience. Therefore, it is necessary to explore alternative methods for capturing videos at a consistent speed, such as using drones, bicycles, or cars. Furthermore, in the constructed virtual world, users are restricted to camera paths of the trajectories of the original recordings, which can limit the sense of free movement. To mitigate this limitation, it may be beneficial to partially incorporate 3D techniques, such as NeRF or Gaussian Splatting, particularly for iconic buildings or landmarks in tourist areas. The difficulty in controlling the avatar was pointed out by participants during the experiment. This issue stems from the fixed camera position, which contrasts with typical third-person games where the camera follows the avatar. Allowing the camera to move freely is expected to enhance the user experience.

# 7 Conclusion

We proposed a system to create realistic virtual worlds of large-scale urban areas and presented evaluation of exploration. Our system builds virtual worlds directly from 360° videos. Our proposed virtual world can be generated automatically from the input videos and does not require any 3D modeling, making it easy for users to create their own environments.

In our user study for the evaluation of 360RVW, we compared 360RVW with GSV and MovieMap under a scenario where users explored specified routes while identifying landmarks. The results indicated that our interface provided a superior sense of exploration because of the proactivity to move via operating an avatar. Furthermore, participants noted that the semantic maps for collision detection felt sufficiently accurate which indicates the potential of this technique for other applications building environments from 2D input data.

**Declarations**

**Competing of interest** The authors have no conflicts of interest.
**Funding** None.
**Authors' contributions** Mizuki Takenawa engaged in research implementation and experiment,

Naoki Sugimoto and Leslie Wöhler helped design of the experiment and implementation of the interface, and Satoshi Ikehata and Kiyoharu Aizawa supervised this research and provided advice and counsel on how to proceed.

**Acknowledgements** This research has been supported by JSPS KAKENHI 23K21677, KAKENHI 25H01164 and CISTI SIP Smart Network System against Natural Disasters.

**Data Availability Statement** Our interface will be available on the web, and data will be made available on reasonable request.